\documentclass{amsart}
\usepackage{stmaryrd}
\usepackage{graphicx}
\usepackage{color}
\usepackage{multirow}
\usepackage{array}
\usepackage{tikz}
\usetikzlibrary{spy}
\usetikzlibrary{backgrounds}
\usetikzlibrary{decorations}
\usetikzlibrary{shapes,snakes}
\usetikzlibrary{shapes.geometric}
\usetikzlibrary{shadows}
\tikzset{
   invisible/.style={opacity=0,text opacity=0},
   visible on/.style={alt={#1{}{invisible}}},
   alt/.code args={<#1>#2#3}{%
     \alt<#1>{\pgfkeysalso{#2}}{\pgfkeysalso{#3}} },}
\tikzset{cross/.style={cross out, draw=black, minimum size=2*(#1-\pgflinewidth), inner sep=0pt, outer sep=0pt},cross/.default={1pt}}
\def\1{1\!{\rm l}}

\newcommand{\minim}[1]{\underset{#1}{\mathrm{min} \ }}
\newcommand{\mubf}{\mbox{\mathversion{bold}{$\mu$}}}
\newcommand{\tbf}{\mbox{\mathversion{bold}{$t$}}}
\newcommand{\Xbf}{\mbox{\mathversion{bold}{$X$}}}
\newcommand{\xbf}{\mbox{\mathversion{bold}{$x$}}}

\def\IF{\mathrm{IF}}
\def\PE{\mathbb{E}}
\def\rset{\mathbb{R}}
\def\rmd{\mathrm{d}}

\definecolor{darkblue}{rgb}{0,0,0.55}
\definecolor{darkgreen}{rgb}{0,0.55,0}
\definecolor{darkred}{rgb}{0.75,0,0}
\definecolor{rougeF}{rgb}{0.65,0.15,0.25}

\makeatletter \@addtoreset{section}{part} \makeatother

\newtheorem{theo}{Théorème}[section]
\newtheorem{prop}[theo]{Proposition}

\newcolumntype{M}[1]{>{\centering\arraybackslash}m{#1}}
\newcolumntype{N}{@{}m{0pt}@{}}

\begin{document}

\title{A breakpoint detection in the mean model with heterogeneous variance on fixed time-intervals}

\author{O. Bock, X. Collilieux, F. Guillamon, E. Lebarbier and C. Pascal}
\email{olivier.bock@ign.fr,Xavier.Collilieux@ign.fr, francois.guillamon@agroparistech.fr,}
\email{emilie.lebarbier@agroparistech.fr, claire.pascal@agroparistech.fr}

\date{\today}

%


\begin{abstract}
This work is motivated by an application for the homogeneization of GNSS-derived IWV (Integrated Water Vapour) series. Indeed, these GPS series are affected by abrupt changes due to equipment changes or environemental effects. The detection and correction of the series from these changes is a crucial step before any use for climate studies. In addition to these abrupt changes, it has been observed in the series a non-stationary of the variability. We propose in this paper a new segmentation model that is a breakpoint detection in the mean model of a Gaussian process with heterogeneous variance on known time-intervals. In this segmentation case, the dynamic programming (DP) algorithm used classically to infer the breakpoints can not be applied anymore. We propose a procedure in two steps: we first estimate robustly the variances and then apply the classical inference by plugging these estimators. The performance of our proposed procedure is assessed through simulation experiments. An application to real GNSS data is presented. 
\end{abstract}

\maketitle

\section{Introduction}
Breakpoint detection aims at detecting abrupt changes, called breakpoints, in the distribution of a signal. Such problems arise in many fields, such as genomics \cite{braun2000multiple,PRL05,cleynen2014comparing,levy2014two}, medical \cite{Lav05}, econometrics \cite{lai2005autoregressive,lavielle,BP03}, geodesy \cite{Gazeaux2013,gazeaux2015joint} or climate \cite{CM2004,mestre2013homer,climat}. This massive number of applications results in an abundant literature on this subject. The motivation of our work comes from the analysis of GNSS-derived Integrated Water Vapour (IWV) series. The IWV plays a significant role in climate studies. However, these series have been affected by abrupt changes due to equipment changes, changes in processing procedure and/or changes in electromagnetic properties of the environment at the measurement site \cite{Vey2009,ning2016uncertainty}. A change in the mean in the signal therefore marks the presence of such an abrupt change. The statistical purpose consists thus in detecting the instants at which the mean changes in the process, that is continuous here. Many approaches have been proposed in the literature about this problem. Among them, we focus on segmentation methods. More precisely, the model of interest will be the following: the signal is supposed to be a realization of an independent Gaussian process whose parameters are affected by an unknown number of changes at unknown times. Two models can be considered, according to the characteristics of the signal that are affected by the changes: it can be either the mean of the signal only (usually called the homoscedastic model) or both the mean and the variance (usually called the heteroscedastic model), as proposed by \cite{PRL05} in a genomic application field or by \cite{gazeaux2015joint} in a geodesic application for the analysis of GPS coordinates series. However, in the GNSS-IWV series, it has been observed a non-stationary of the variance due to increased variability of IWV in summer. Inspection of the annual variation of the series shows that a monthly sampling of the variance will be adequate. Consequently, the two above models will fail. \\
The model we propose in this work is thus a segmentation in the mean of a Gaussian process model with heterogeneous variances in the sense that the stationarity time-intervals of the variance are fixed (the months for the application). \\

It is now well known in segmentation framework that segmentation raises algorithmic issues due to the discrete nature of the breakpoint parameters. Indeed, the inference of these parameters requires to visit the whole segmentation space, which is prohibitive in terms of computational time when the visit is performed in a naive way. The Dynamic Programming (DP) algorithm (introduced by \cite{Bellman} and used for the first time in segmentation by \cite{Auger89}) and, recently its pruned versions \cite{killick2012,Rigaill2015,maidstone2014}, is the only efficient algorithm that retrieves the exact solution (i.e. the optimal segmentation according to the $\log$-likelihood or least-square contrasts for example) in a faster way. However this algorithm can only be used if the quantity to be optimized is segment-additive (see for example \cite{BP03} or \cite{Lav05}). In other words, a sufficient condition to satisfy this assumption is the fact that the segments are not linked both in terms of observations (i.e. independence) and parameters (i.e. no common parameters). In our case, the both stationary time-intervals of the means and the variances do not coincide. Two problems will appear: first the estimators of these two parameters will be linked and then we have no hope that DP can be applied. In order to circumvent this problem and retain the use of DP, we consider the same inference strategy as in \cite{CLLR2015} or \cite{Cleynen_Robin2014} which consists in a two-step procedure: we first estimate the 'nuisance' parameters (here the variances) and then we apply the classical inference procedure by plugging these estimators. \\

The problem is thus reduced to the estimation of the variance parameter in a series with changes in the mean. Due to the presence of breakpoints in the series, the classical estimators for the variance will fail. Here, we follow the same idea as in \cite{CLLR2015} who proposed a robust estimator of the autocorrelation parameter for estimating breakpoints in the mean of an AR(1) process. Briefly speaking, instead of using the raw series, the idea is to work with the differenciated series that is then a zero-mean Gaussian process except at the position of the breakpoints. These points can be then seen as outliers and a robust approach can be used to obtain a good estimator of the scale parameter, as \cite{CR} proposed. We adapt in particular this estimator to our case for which, using the results of \cite{levy2011robust}, we obtain asymptotic properties. \\ 
For the second step of the inference, if DP can be applied to obtain the best segmentation of the series in a given number of segments, the question arises of the choice of this number. This question has been widely investigated. In this paper, we propose to adapt the criteria proposed by \cite{Lav05}, \cite{L05} and \cite{ZhS07}. \\

This paper is organized as follows: Section \ref{sec:model} presents the proposed segmentation model, describes the algorithmic issue for the inference and gives the outline of the proposed inference strategy. The details of this strategy are given in Section \ref{sec:inference}. More precisely, the robust estimator of the variance and the different model selection criteria for choosing the number of segments are given. A simulation study is performed in Section \ref{sec:simulation} and Section \ref{sec:application} is dedicated to an application of our method on GNSS-derived IWV series. 

\section{Model and inference issue}  \label{sec:model}
\subsection{Model}

We observe a series $y=\{y_t\}_{t=1,\ldots,n}$ modeled by a Gaussian independent random process $Y=\{Y_t\}_{t=1\ldots,n}$ such that 
\begin{description}
\item[$\star$] the mean of $Y$ is affected by $K-1$ abrupt changes at some unknown instants, called breakpoints, $0=t_0<t_1<\ldots<t_{K-1}<t_K=n$ and is constant between two breakpoints or within the interval $I_{k}^{\text{mean}} =
\llbracket t_{k-1}+1, t_{k} \rrbracket$, denoted segment, and
\item[$\star$] the variance of $Y$ is also subject to known $J-1$ changes, i.e. the variance is constant within each interval $I^{\text{var}}_j$ and different from one to another.
\end{description}
The model is thus the following:
\begin{equation} \label{model:m1}
Y_{t} \sim \mathcal{N}(\mu_k,\sigma_{\text{j}}^2) \ \ \text{$\forall t \in I_{k}^{\text{mean}} \cap I^{\text{var}}_j$,}  
\end{equation}
for $k=1,\ldots,K$ with $K$ is the number of segments or intervals $I_{k}^{\text{mean}}$ and for $j=1,\ldots,J$ with $J$ is the number of intervals $I^{\text{var}}_j$. Contrary to the heteroscedastic model, the intervals $I^{\text{var}}_j$ and $I_{k}^{\text{mean}}$ are not assumed to be the same. 


\subsection{Segmentation inference: an algorithmic issue}

Parameter inference in model \eqref{model:m1} amounts to estimating the number of segments $K$, the breakpoints $\tbf = (t_k)_{k=1,\ldots,K-1}$ and the distribution parameters, i.e. the means $\mubf=(\mu_k)_{k=1,\ldots,K}$ and the variances ${\bf{\sigma}}^2=(\sigma^2_{j})_{j=1,\ldots,J}$. To this end, we use a (penalized) maximum-likelihood framework and proceed as classically in segmentation inference in three steps: (i) estimate the distribution parameters, the breakpoints and their number being fixed, (ii) estimate the breakpoints for a fixed $K$ and (iii) choose the number of segments $K$. \\

The $\log$-likelihood of model \eqref{model:m1} is 
\begin{equation} \label{eq:loglik_m1}
\log p(y; \tbf, \mubf, {\bf{\sigma}})= - \frac{n}{2} \log{(2 \pi)} - \sum_{j=1}^J \frac{n_{j} }{2}  \log{(\sigma^2_j)}-\frac{1}{2} SSwg_K(\tbf,\mubf,{\bf{\sigma}}^2),
\end{equation}
where
\begin{equation} \label{eq:SSwg}
SSwg_K(\tbf,\mubf,{\bf{\sigma}}^2)=\sum_{k=1}^K \sum_{j=1}^J \sum_{t \in I_{k}^{\text{mean}} \cap I^{\text{var}}_j} \frac {(y_t-\mu_k)^2}{\sigma^2_j},
\end{equation}
and $n_j$ is the length of interval $I^{\text{var}}_j$. Recall that in the segmentation framework, it is now well known that the step (ii) leads to a discret optimization problem and that the only efficient algorithm that retrieves the solution (exact solution in a fast way) is the Dynamic Programming algorithm (DP). This algorithm can be applied under the constraint that the quantity to be optimized is additive with respect to the segments or intervals $I_{k}^{\text{mean}}$ (see for example \cite{BP03}, \cite{PRL05} or \cite{Lav05}). Here the optimization problem for breakpoint estimation is 
\begin{equation*}
\minim{\tbf \in \mathcal{M}_{K,n} } \minim{\mubf \in \mathbb{R}^K} \minim{\bf{\sigma} \in {\mathbb{R}^{+}}^J} -\log p(y; \tbf, \mubf, {\bf{\sigma}})= \minim{\tbf \in \mathcal{M}_{K,n}}  -\log p(y; \tbf, \widehat{\mubf}, \widehat{\bf{\sigma}}),
\end{equation*}
where $\log p(y; \tbf, \mubf, \bf{\sigma})$ is given in \eqref{eq:loglik_m1} and $\mathcal{M}_{K,n}=\{(t_1,\ldots,t_{K-1}) \in \mathbb{N}^{K-1}, 0=t_0<t_1<\ldots,t_{K-1}<t_K=n\}$ is the set of all possible partitions in $K$ segments of the grid $\llbracket 1,n \rrbracket$. The carriers of the mean parameters and the variance parameters being not the same, $I_{k}^{\text{mean}}$ for $\mu_k$ and $I^{\text{var}}_j$ for $\sigma^2_j$, two problems appear: first the estimators of these two parameters are linked, as we observe on their expressions:
\begin{equation} \label{eq:esti_moy_var_m1}
\widehat{\mu}_k = \frac{\sum_{j=1}^J \sum_{t \in I_{k}^{\text{mean}} \cap I^{\text{var}}_j} \frac{Y_t}{\widehat{\sigma}_{j}^2 }} { \sum_{j=1}^J  \sum_{t \in I_{k}^{\text{mean}} \cap I^{\text{var}}_j} \frac{1}{\widehat{\sigma}_{j}^2}} \ \ , \ \ \widehat{\sigma}_j^2 = \frac{1}{n_j} \sum_{k=1}^K \sum_{t \in I_{k}^{\text{mean}} \cap I^{\text{var}}_j} (Y_t - \widehat{\mu}_k)^2.
\end{equation}
Then we have no hope that $-\log p(y; \tbf, \widehat{\mubf}, \widehat{{\bf{\sigma}}})$ will be segment-additive so DP can not be used to estimate the breakpoints.

\section{Inference procedure} \label{sec:inference}
In order to keep possible the use of DP, we consider the same strategy as proposed by \cite{CLLR2015} or \cite{Cleynen_Robin2014} which consists in \begin{description}
\item[(1)] estimating the variance parameters (see Section \ref{sec:robust}), the estimators are denoted $\tilde{\sigma}^2_j$,
\item[(2)] using the classical inference with 'known' variances. In this case,
\begin{description}
\item[$\star$] the mean estimators are the same as \eqref{eq:esti_moy_var_m1} where $\widehat{\sigma}_{j}^2$ is replaced by $\tilde{\sigma}^2_j$,
\item[$\star$] the optimization problem for breakpoint estimation is 
\begin{eqnarray} \label{eq:Contraste_Pour_DP}
\minim{\tbf \in \mathcal{M}_{K,n} }-\log p(y; \tbf, \widehat{\mubf},\widehat{\sigma}) &=& \minim{\tbf \in \mathcal{M}_{K,n} } \sum_{k=1}^K \sum_{j=1}^J \sum_{t \in I_{k}^{\text{mean}} \cap I^{\text{var}}_j} \frac{(y_t-\widehat{\mu}_k)^2}{\tilde{\sigma}^2_j} \\
&=& \minim{\tbf \in \mathcal{M}_{K,n} }SSwg_K(\tbf,\widehat{\mubf},\tilde{\sigma}^2) \nonumber \\
&=& SSwg_K(\widehat{\tbf},\widehat{\mubf},\tilde{\sigma}^2),\nonumber 
\end{eqnarray}
and DP applied. 
\item[$\star$] the number of segments $K$ is chosen according to a model selection strategy which consists in maximizing a penalized $\log$-likelihood (see Section \ref{sec:model_selection}). 
\end{description}
\end{description}

\subsection{A robust estimator of the scale parameters in presence of breakpoints}  \label{sec:robust}

For a sake of simplicity, let us consider that the variance of the process $Y$ is homogeneous, i.e. in model \eqref{model:m1} we have $J=1$, $\sigma^2_j=\sigma^2$ whatever $j$ (i.e. $I^{\text{var}}_j=\llbracket 1,n \rrbracket$) and the purpose is to estimate $\sigma$. Since we need to estimate it in a series with breakpoints, classical estimators failed. The objective is to provide a robust (faced to the presence of breakpoints) estimator of $\sigma$. Following \cite{CLLR2015}, the idea is to work on the differentiated series $X_t=(Y_{t+1} - Y_t)_t$ since the means of this novel series is equal to $0$ except at the breakpoint positions (only $K-1$ ($K \ll n$) differences are non-centered). These latter breakpoints can then be seen as outliers and robust approaches can be used to estimate $\sigma$. \cite{CR} proposed a robust estimator of the scale parameter of an independent Gaussian stationary process $\Xbf$ that is proportional to the first quartile of the $n^2$ differences $\left\lbrace \left| X_i - X_j\right| ; \; 1\leq i < j\leq n \right \rbrace$, i.e.
\begin{equation} \label{eq:robust_sigma_CR} 
Q_{CR,n}(\Xbf) =  c_{Q} \left\lbrace \left| X_i - X_j\right| ; \; 1\leq i < j\leq n \right \rbrace_{\left(\left\lceil \frac{1}{4} C_n^2 \right\rceil\right)}, 
\end{equation}
with 
\begin{equation} \label{eq:constante_c}
 c_{Q} = \frac{1}{\sqrt{2}\Phi^{-1}\left(\frac{5}{8}\right)} \approx 2.2191,
\end{equation}
to ensure the consistency of the estimator, and where $\Phi$ denotes the cumulative distribution function of a standard Gaussian random variable. The asymptotic properties of this estimator have been studied by \cite{levy2011robust} for Gaussian stationary processes with either short-range or long-range dependence. \\
Using this estimator, the robust estimator $\sigma$ we proposed in our context is given in Proposition \ref{prop:robust} for which asymptotic properties are obtained. 

\begin{prop}\label{prop:robust}
Let $(Y_t)_t$ and $(E_t)_t$ such that $Y_t=\mu_k+E_t$ if $t \in I_{k} =\llbracket t_{k-1}+1, t_{k} \rrbracket$ for $k=1,\ldots,K$ where $(E_t)_t$ are i.i.d centered Gaussian with variance $\sigma^2$ and let further assume that $Y_0 \sim \mathcal{N}(\mu_1,\sigma^2)$. Let denote $\Xbf=(X_t)_{t=0,\ldots,n-1}=(Y_{t+1}-Y_{t})_{t=0,\ldots,n-1}$ and $(\nu_t)_{t=0,\ldots,n-1}=(E_{t+1}-E_{t})_{t=0,\ldots,n-1}$. Let 
\begin{equation} \label{eq:esti_sigma}
\tilde{\sigma}_{n}=Q_{n}(\Xbf)=  c_{Q} \ \frac{Q_{CR,n}(\Xbf)}{\sqrt{2}},
\end{equation}
where $Q_{CR,n}(\Xbf)$ and $c_{Q}$ are given in \eqref{eq:robust_sigma_CR} and \eqref{eq:constante_c} respectively. Then, ${Q}_n$ satisfies the following Central Limit Theorem
\begin{equation*}
\sqrt{n}({Q}_n(\Xbf) - \sigma) \stackrel{d}{\longrightarrow}\mathcal{N}(0,\sigma^{\prime 2})\;, \textrm{ as } n\to\infty\;,
\end{equation*}
where 
\begin{eqnarray*}
{\sigma}^{\prime 2} & = & \sigma \PE[\IF^2 \left( \nu_{0}/{\sqrt{2} \sigma}\right)] +2\sigma  \sum_{h \geq 1}\PE\left[\IF (\nu_{0}/{\sqrt{2} \sigma}) \IF (\nu_{h}/{\sqrt{2} \sigma}) \right]\; , \\
\IF(x) & = & c_{Q}\left(\frac{1/4-\Phi(x+1/c_{Q}) +\Phi(x-1/c_{Q})}{\int_{\rset} \phi(y)\phi(y+1/c_{Q,\Phi})\rmd y}\right),
\end{eqnarray*}
and where $\Phi$ and $\phi$ denote the cumulative distribution function and the probability distribution function of a standard Gaussian random variable, respectively.
\end{prop}

Proof: since the proposed estimator is proportional to the CR's one \eqref{eq:robust_sigma_CR}, the asymptotic result is simply obtained using the results obtained by \cite{levy2011robust}: Theorem 2 is applied on $(\nu_t)_t$ with $\gamma_{\nu}(0)=2 \sigma^2$ and since $\sum_{h \geq 1} |\gamma_{\nu}(h)| < \infty$.\\

Note that by working on the differenciated series $(X_t)_{t=0,\ldots,n-1}=(Y_{t+1}-Y_{t})_{t=0,\ldots,n-1}$, the dependence is lost but it is of short-range and remark that $X_t$ is a Gaussian process with variance $2 \sigma^2$ (that explained the normalization by $\sqrt{2}$ in \eqref{eq:esti_sigma}). \\

Come back to our segmentation model \eqref{model:m1} and using Proposition \ref{prop:robust}, the proposed estimator of $\sigma_j$ is 
\begin{equation}\label{eq:esti_robust__sigmaj}
\tilde{\sigma}_{j,n}= Q_{n}(\Xbf^{(j)}),
\end{equation}
where $\Xbf^{(j)}=(X^{(j)}_t)_t=(Y_{t+1}-Y_{t})_{t \in I^{\text{var}}_j}$. We note $\tilde{\sigma}_n=(\tilde{\sigma}_{j,n})_j$.

\subsection{Selecting the number of segments} \label{sec:model_selection}

In order to select the number of segments $K$, we consider three criteria proposed by \cite{Lav05}, \cite{L05} and \cite{ZhS07}. We use these criteria forgetting the fact that $\sigma^2_j$ has been estimated in a first step. The two first criteria are penalized contrast criteria which differ from the form of the penalty and depend on constants to be calibrated contrary to the last one. The penalty proposed by \cite{Lav05}, denoted Lav, depends on the number of parameters in a model with dimension $K$ (i.e. a segmentation with $K$ segments) denoted $D_K$. It is defined as follows:
\begin{equation} \label{eq:critere_Marc}
\text{Lav} (K)= SSwg_K(\widehat{\tbf},\widehat{\mubf},\tilde{\sigma}^2)+\beta D_K,
\end{equation}
where $SSwg_K(\tbf, \mubf,\sigma^2)$ is the sum of squares given in \eqref{eq:SSwg}. $D_K=K$, the $K$ means. The constant $\beta$ is chosen using an
adaptative method which involves a threshold $s$, taken in the simulation study and the applications to $s=0.7$ as suggested by \cite{Lav05}. \\
Applying the works of \cite{BM2001} in the segmentation context, \cite{L05} proposed a more complex penalty in which, in addition to $D_K$, the number of possible segmentations with $K$ segments (that is $\binom{n-1}{K-1}$) is taken into account for. This criterion is denoted BM and is defined as follows:
\begin{equation} \label{eq:critere_Leb}
\text{BM} (K)=SSwg_K(\widehat{\tbf},\widehat{\mubf},\tilde{\sigma}^2)+\alpha \left[ 5 D_K +2 K \log \left
(\frac{n}{K} \right) \right].
\end{equation}
This penalty also depends on a constant $\alpha$ which can be
calibrated in practice using the slope heuristic method proposed in \cite{Arl_Mas-pente}. More pecisely, there exists two algorithms based on this heuristic: the dimension jump algorithm and the data-driven slope estimation algorithm. We use for the simulations and the application the package R \texttt{capushe} \cite{capushe} and denote BM1 and BM2 the criteria BM where the constant is calibrated using these two algorithms respectively. Note that these criteria have to be minimized. \\
The last criterion is a modified version of the classical BIC criterion \cite{Schwarz1978} adapted by \cite{ZhS07} to the segmentation in the mean with homogeneous variance framework, and so-called mBIC. Two versions are derived depending on the knowledge or not of the variance. Here, we considered the one for which the variance is supposed to be known, denoted mBIC,
\begin{equation} \label{eq:critere_mBIC1}
\text{mBIC} (K)=- \frac{1}{2} SSwg_K(\widehat{\tbf},\widehat{\mubf},\tilde{\sigma}^2) -\frac{1}{2} \sum_{k=1}^K \log {(\widehat{n}_k)}+\left (\frac{3}{2}-K \right ) \log{(n)},
\end{equation}
where $\widehat{n}_k = \widehat{t}_k - \widehat{t}_{k-1}$ is the length of the $k$th segment of the best segmentation with $K$ segments (i.e. of $\widehat{\tbf}$). Note that this criterion has to be maximized.

\section{Simulation study} \label{sec:simulation}

In order to assess the performance of our procedure, we conduct the simulation study described below. Note that we indiced the true parameters by $^{\star}$.

\subsection{Simulation design and quality criteria}

We use a similar design as in \cite{CLLR2015} for the segmentation parameters (breakpoint locations and means) and mimic our motivation application in the sense that the series include several years and the variance time-intervals correspond to the months. We consider series of length $n \in \{200, 800\}$ with $4$ years of $n/4$ points each and $2$ months by year with standard deviation $\sigma_1^{\star}$ and $\sigma_2^{\star}$ respectively. $\sigma_1^{\star}$ is fixed to $0.5$ and $\sigma_2^{\star}$ varies from $0.1$ to $1.5$ by step of $0.2$. The series are affected by $6$ breakpoints ($K^{\star}=7$, the true number of segments) located at positions $\tbf^\star=(27, 38, 88, 111, 150, 183)$ for $n=200$ and $4 \times \tbf^\star$ for $n=800$. The mean within each segment alternates between 0 and 1, starting with $\mu_1 = 0$. Each configuration is simulated $100$ times. \\

Moreover, in order to exhibit the need of a new segmentation model for our motivated application, we compare our segmentation method, called  MFixedHetero, with the two more classical segmentation models (see for example \cite{PRL05}) with 
\begin{description}
\item[$\star$] changes in the mean with homogeneous variance, called MHomo: $Y_t \ ind. \sim \mathcal{N}(\mu_k,\sigma^2)$ if $t \in I_{k} =\llbracket t_{k-1}+1, t_{k} \rrbracket$,

\item[$\star$] changes in both the mean and the variance, called MHetero: $Y_t \ ind. \sim \mathcal{N}(\mu_k,\sigma_k^2)$ if $t \in I_{k} =\llbracket t_{k-1}+1, t_{k} \rrbracket$.\\
\end{description}

In order to evaluate the performance of our proposed method, we use the following criteria:
\begin{description}
\item[$\star$] the difference between the estimated standard deviation and the true one, $\tilde{\sigma}_{\bullet,n}-\sigma_{\bullet}^\star$,
\item[$\star$] the difference between the estimated number of segments and the true one, $\hat{K}-K^{\star}$,
\item[$\star$] the two components of the Hausdorff distance $d_1(\tbf^{\star}, \hat{\tbf})$ and $d_2 (\tbf^{\star}, \hat{\tbf})$ where 
$$d_1(a,b)=\max_b \min_a |a-b|,$$
and $d_2(a,b)=d_1(b,a)$, in order to study the quality of the estimation of the breakpoint locations. A perfect segmentation results in both null $d_1$ and $d_2$. Under-segmentation results in a small $d_1$ and a large $d_2$, provided that the estimated breakpoints are correctly located. \\
\end{description}

\subsection{Results}

\begin{figure}[h!]
\centering
\begin{tikzpicture}
\node at (0,0) {\includegraphics[height=5cm,width=7cm]{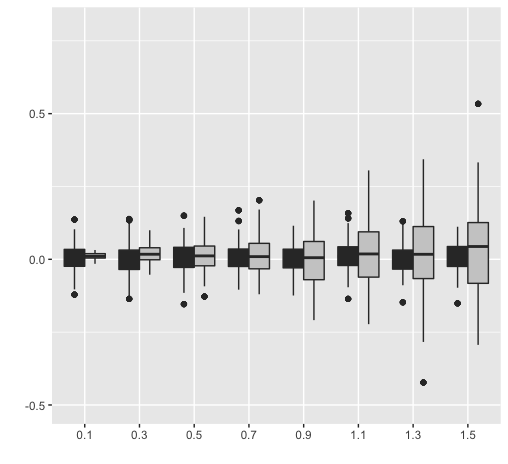}};
\node at (7.5,0) {\includegraphics[height=5cm,width=7cm]{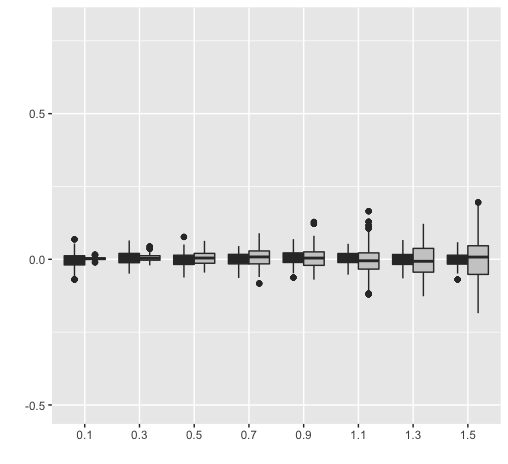}};
 \end{tikzpicture}
\caption{Boxplots of $\tilde{\sigma}_{1,n}-\sigma_1^\star$ in black and $\tilde{\sigma}_{2,n}-\sigma_2^\star$ in grey for different values of $\sigma_2^\star$ with $n=200$ (left) and $n=800$ (right).} \label{fig:variance}
\end{figure}

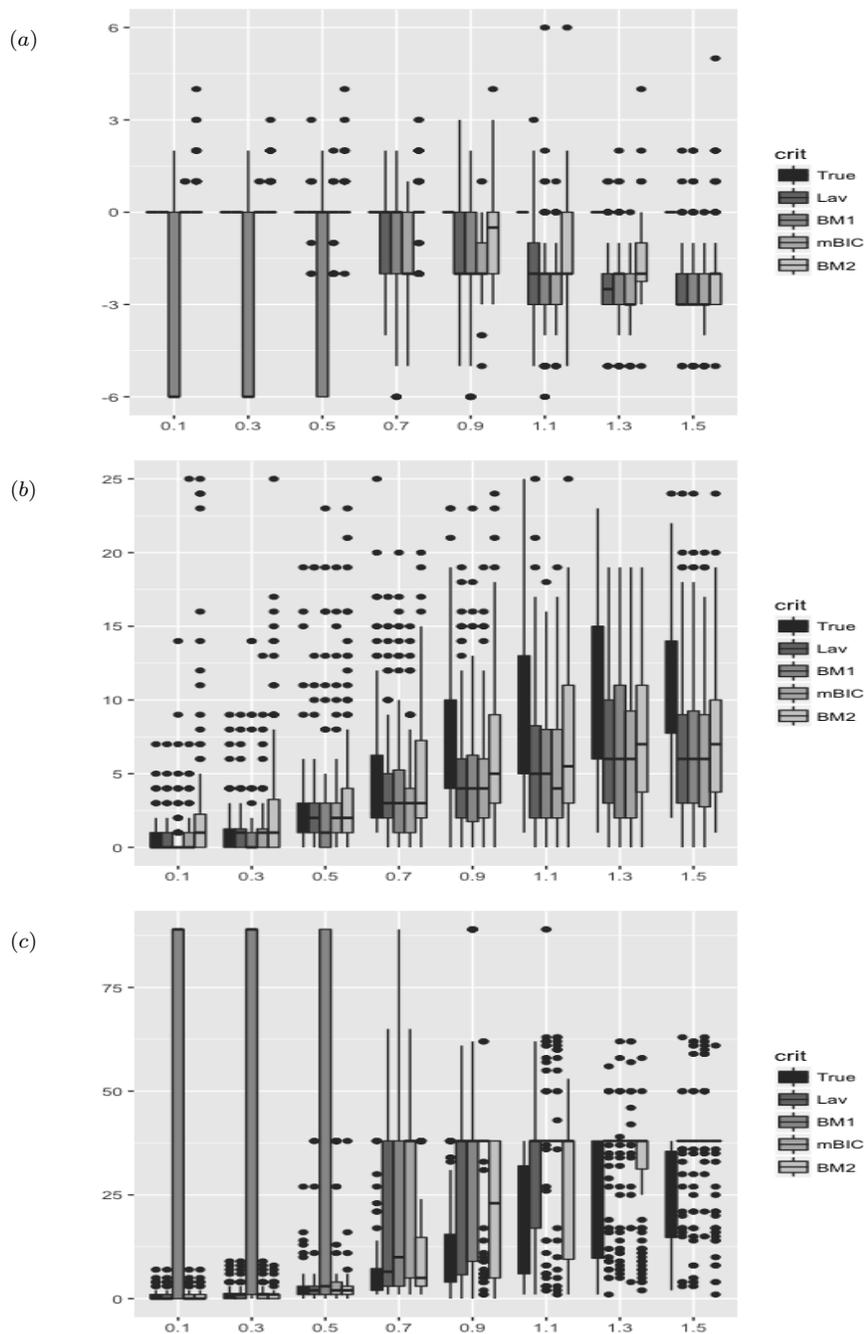
\begin{figure}[h!]
\centering
\begin{tikzpicture}
\node at (0,0) {\includegraphics[height=6cm,width=11cm]{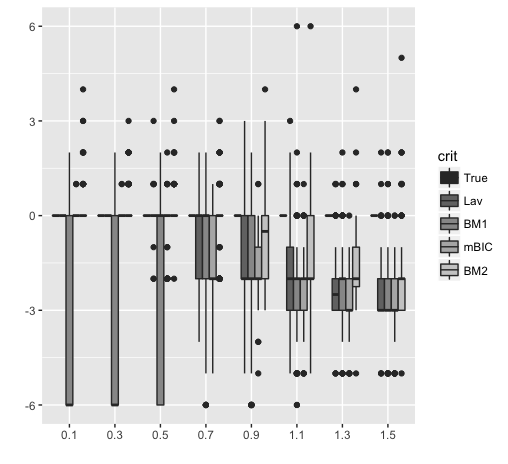}};
\node at (0,-6) {\includegraphics[height=6cm,width=11cm]{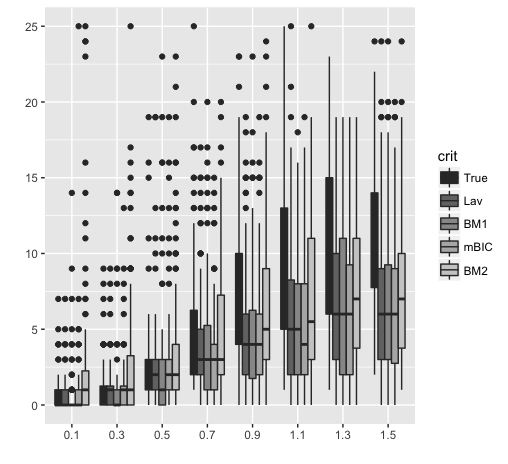}};
\node at (0,-12) {\includegraphics[height=6cm,width=11cm]{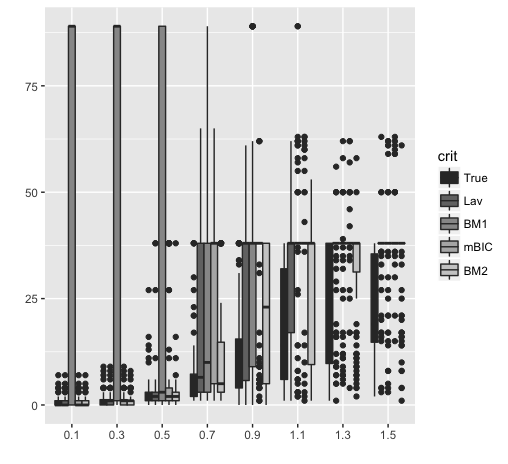}};
\node at (-6,2.5) {\footnotesize{$(a)$}};
\node at (-6,-3.5) {\footnotesize{$(b)$}};
\node at (-6,-9.5) {\footnotesize{$(c)$}};
 \end{tikzpicture}
\caption{Boxplots of (a) $\hat{K}-K^\star$, (b) the first component of the Hausdorff distance ($d_1$) and (c) the second component of the Hausdorff distance ($d_2$) for $n=200$ and for different values of $\sigma^\star_2$, obtained for MFixedHetero and the different model selection criteria. } \label{fig:Study2}
\end{figure}

\begin{figure}[h!]
\centering
\begin{tikzpicture}
\node at (0,0) {\includegraphics[height=6cm,width=11cm]{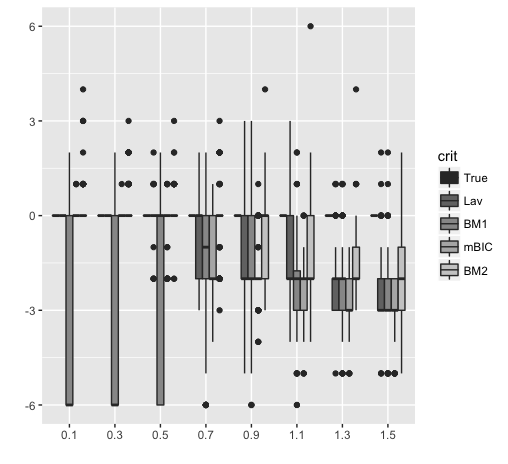}};
 \end{tikzpicture}
\caption{Boxplots of $\hat{K}-K^\star$ for $n=200$ and for different values of $\sigma^\star_2$, obtained for MFixedHetero with the true values of $\sigma_1^\star$ and $\sigma_2^\star$ and the different model selection criteria.} \label{fig:Study2_TRUE}
\end{figure}

\begin{figure}[h!]
\centering
\begin{tikzpicture}
\node at (0,0) {\includegraphics[height=4cm,width=5cm]{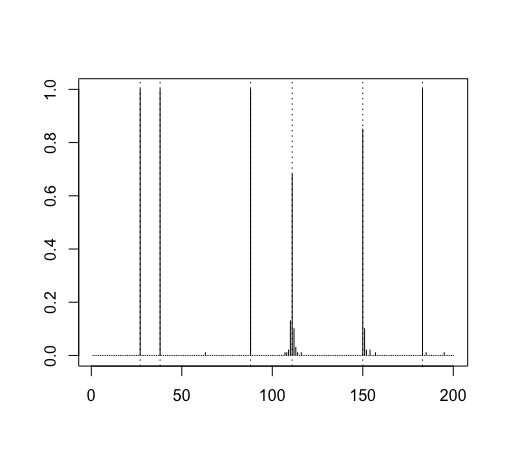}};
\node at (5,0) {\includegraphics[height=4cm,width=5cm]{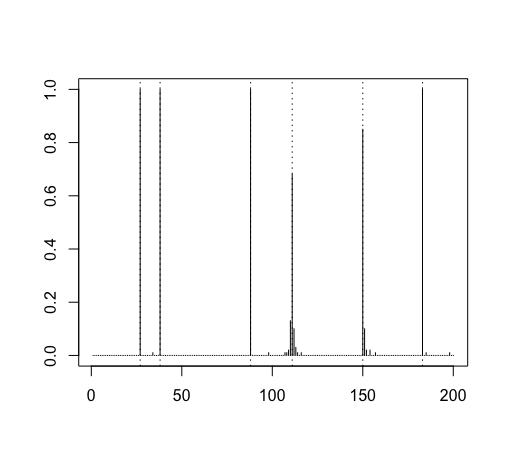}};
\node at (10,0) {\includegraphics[height=4cm,width=5cm]{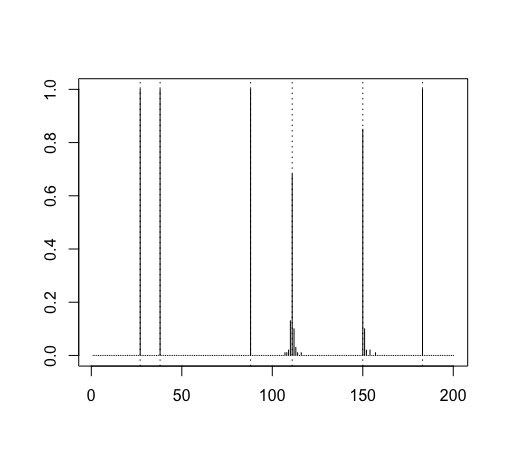}};
\node at (0,-4) {\includegraphics[height=4cm,width=5cm]{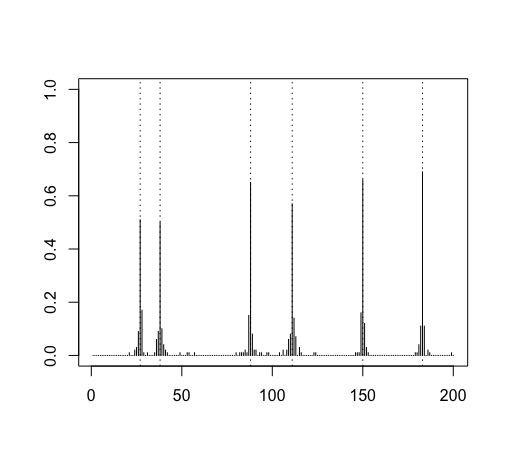}};
\node at (5,-4) {\includegraphics[height=4cm,width=5cm]{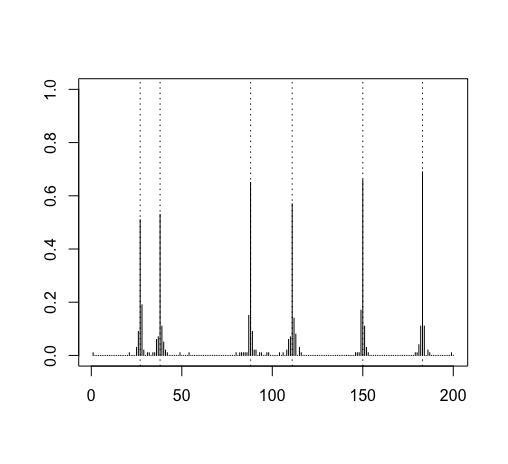}};
\node at (10,-4) {\includegraphics[height=4cm,width=5cm]{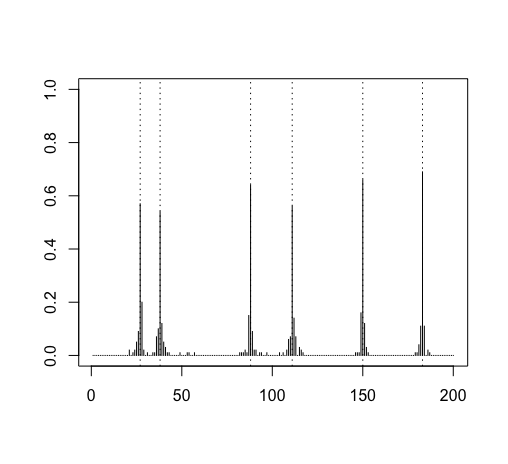}};
\node at (0,-8) {\includegraphics[height=4cm,width=5cm]{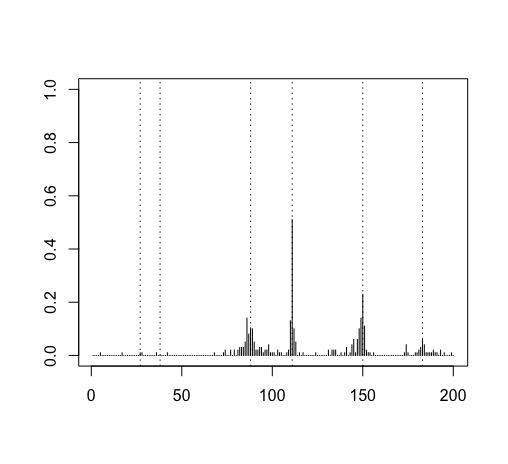}};
\node at (5,-8) {\includegraphics[height=4cm,width=5cm]{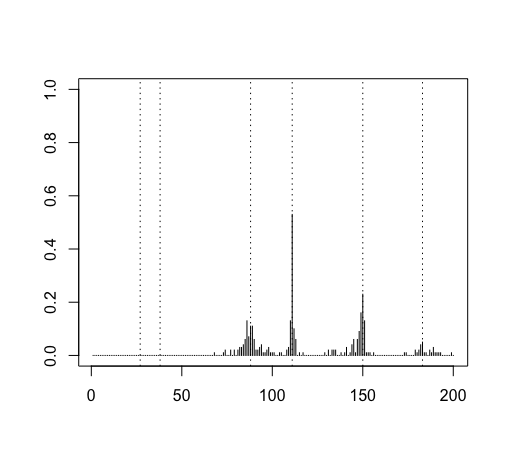}};
\node at (10,-8) {\includegraphics[height=4cm,width=5cm]{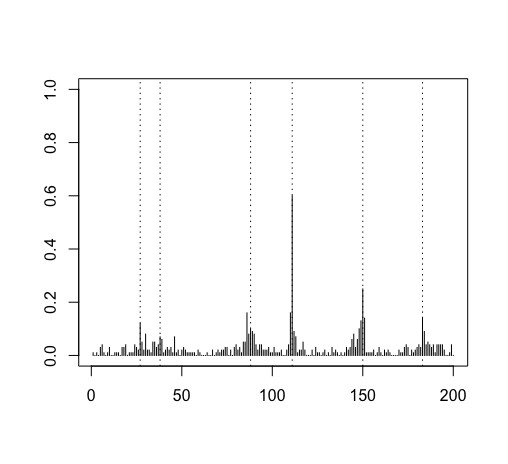}};
\node at (-2.6,1.3) {\footnotesize{$(a)$}};
\node at (-2.6,-2.7) {\footnotesize{$(b)$}};
\node at (-2.6,-6.7) {\footnotesize{$(c)$}};
 \end{tikzpicture}
\caption{Frequencies of each possible breakpoint for MFixedHetero when the number of segments is selected with mBIC and the variances are estimated (left), when the number of segments is selected with mBIC and the variances are the true values (middle) and when the number of segments is true ($K=7$) and the variances are estimated (right), with $n=200$. The value of  $\sigma_2^\star$ is fixed to $0.1$ (a), $0.5$ (b) and  $1.5$ (c). The dotted lines correspond to the true breakpoint locations.} \label{fig:Hist_rupt_mod3}
\end{figure}

\paragraph{\bf {Estimation of $\sigma_1^{\star}$ and $\sigma_2^{\star}$.}} Figure \ref{fig:variance} presents the proposed estimator for the two variances $\sigma_1^\star$ and $\sigma_2^\star$. We observe that it performs well to estimate the variances resulting in a similar performance in terms of segmentation estimation (see Figures \ref{fig:Study2}(a) and \ref{fig:Study2_TRUE} for the selection of $K$, and Figure \ref{fig:Hist_rupt_mod3} for the locations of the breakpoints when the variances are estimated or fixed to the true values). We can also note that the accuracy of the variance estimations increases with the length of the series $n$. \\

\paragraph{\bf {Segmentation estimation for MFixedHetero.}} Only the results for $n=200$ are presented here, the results for $n=800$ leading to the same conclusions. \\

Figure \ref{fig:Study2}(a) compares the estimated number of segments obtained with the considered model selection criteria for different noise levels of $\sigma_2^\star$. The two components of the Haussdorff distance ($d_1$ and $d_2$) calculated on the obtained segmentations are plotted in Figures \ref{fig:Study2}(b) and \ref{fig:Study2}(c) respectively. These distances are also computed for the optimal segmentations with the true number of segments (on the same figures). In addition, the histograms of breakpoint locations are given in Figure \ref{fig:Hist_rupt_mod3} for three values of $\sigma^\star_2$ when the number of segments is selected using mBIC or fixed to the true value and when the variances are estimated or not (the other criteria giving similar results). \\

First recall that in a segmentation in the mean context, it has been observed that when the noise is small, the detection problem is easy and the procedure detects the true breakpoints. However, when the problem gets difficult (large variance), the procedure tends to underestimate the  number of segments in order to avoid the detection of false breakpoints (see for example \cite{CLLR2015}). In our simulation design, among the six breakpoints, four belong to an interval with variance $\sigma_2^{2 \star}$, the fourth one, $t^\star_4$, belongs to an interval with variance $\sigma_1^{2 \star}$ and the fifth one $t^\star_5$ corresponds to both a change in the mean and the variance. We thus observe that our procedure performs as expected and whatever the model selection criteria. First, the variance $\sigma_2^{2 \star}$ does not alter the detection of the breakpoint $t^\star_4$. When $\sigma_2^{2 \star}$ is small, all the true breakpoints are recovered with a less of accuracy for $t^\star_4$ and $t^\star_5$, and when $\sigma_2^{2 \star}$ becomes large, the procedure tends to underestimate the number of segments with estimated breakpoints that are close to the true ones ($d_1$ smaller compared to the segmentations with the true number of segments). For a very high value of $\sigma_2^{2 \star}$, almost only $t^\star_4$ is detected. We can also observe that our method performs as well as when the variances are known. Moreover, even if the different criteria for selecting the number of segments show a global same behaviour, there exist some slight differences: BM1 fails when the detection problem is very easy due to the calibration heuristic, BM2 tends to detect a little more number of segments compared to the other criteria when the variance is large.\\

\paragraph{\bf {Comparison with models MHomo and MHetero.}}

\begin{figure}[h!]
\centering
\begin{tikzpicture}
\node at (0,0) {\includegraphics[height=5cm,width=8cm]{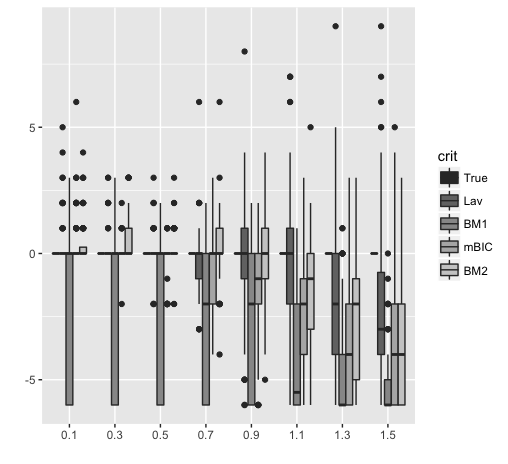}};
\node at (8,0) {\includegraphics[height=5cm,width=8cm]{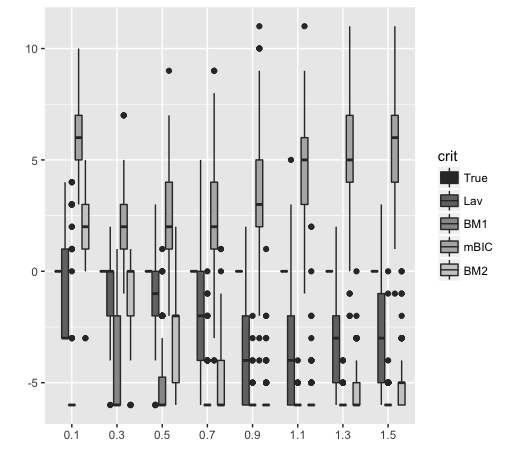}};
 \end{tikzpicture}
\caption{Boxplots of $\hat{K}-K^\star$ for $n=200$ and different values of $\sigma_2^\star$ with MHomo (left) and MHetero (right) using the different model selection criteria.} \label{fig:SelectedKOtherModels}
\end{figure}

Figure \ref{fig:SelectedKOtherModels} displays the boxplots of the number of segments selected by Lav, BM and mBIC for models MHomo and MHetero and Figures \ref{fig:Hist_rupt_mod1} and \ref{fig:Hist_rupt_mod2} give the histograms of the breakpoint locations for the different model selection criteria and three values of $\sigma_2^\star$, obtained with MHomo and MHetero respectively. Note that the imposed changes of variance are located at the positions $25, 50, 75, 100, 125, 150, 175$ and the true breakpoints at $27, 38, 88, 111, 150, 183$. \\   
Logically MHomo, MHetero and MFixedHetero lead to close performances in terms of segmentation when the two variances are close, even if for MHetero we can observe an overestimation by mBIC (Figure \ref{fig:Hist_rupt_mod2} (b-left) or Figure \ref{fig:SelectedKOtherModels} (right)) and a less of accuracy with Lav and BM2. With model MHetero, as expected, the changes in the variance are also detected, with more difficulty compared to the detection of the changes in the mean. This explained the overestimation of the estimated number of segments. This is more marked with mBIC. Model MHomo behaves similarly as model MFixedHetero, except when the variance is too large (Figure \ref{fig:Hist_rupt_mod1} (c)). In this latter case, MFixedHetero can be able to detect the fourth breakpoint $t^\star_4$ contrary to MHomo for which the estimated standard deviation is larger than $0.5$ in the corresponding interval ($1.27$ in average).

\begin{figure}[h!]
\centering
\begin{tikzpicture}
\node at (0,0) {\includegraphics[height=4cm,width=5cm]{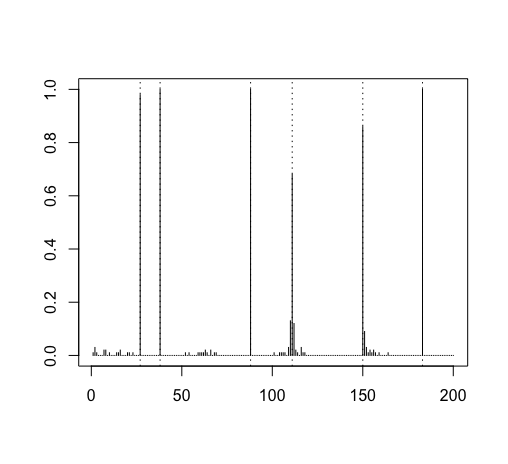}};
\node at (5,0) {\includegraphics[height=4cm,width=5cm]{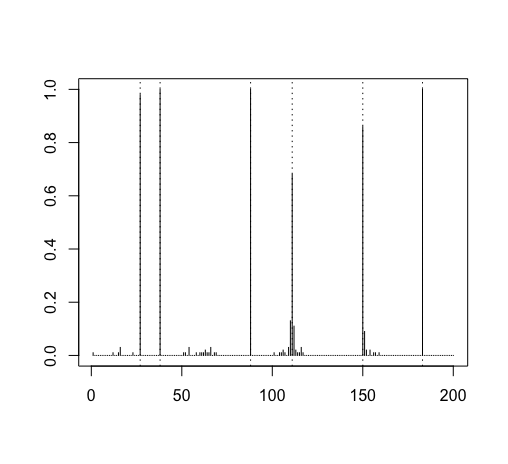}};
\node at (10,0) {\includegraphics[height=4cm,width=5cm]{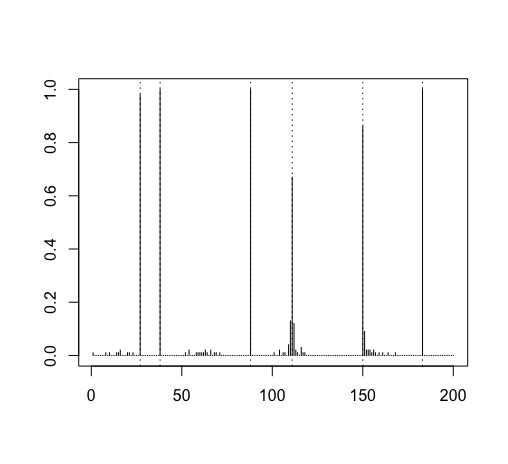}};
\node at (0,-4) {\includegraphics[height=4cm,width=5cm]{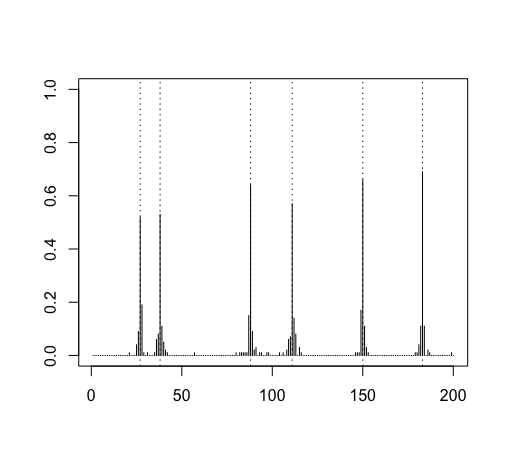}};
\node at (5,-4) {\includegraphics[height=4cm,width=5cm]{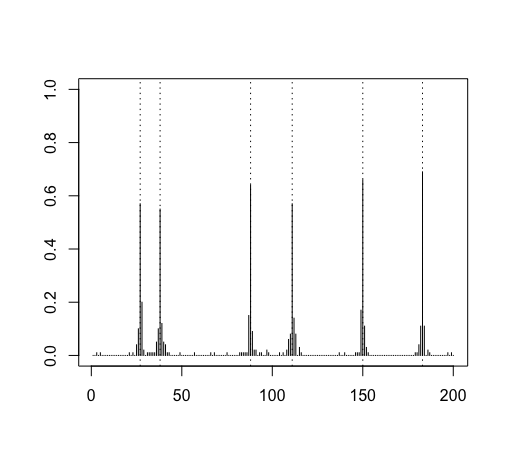}};
\node at (10,-4) {\includegraphics[height=4cm,width=5cm]{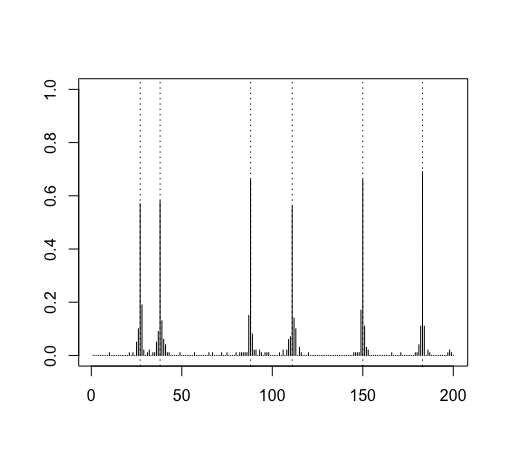}};
\node at (0,-8) {\includegraphics[height=4cm,width=5cm]{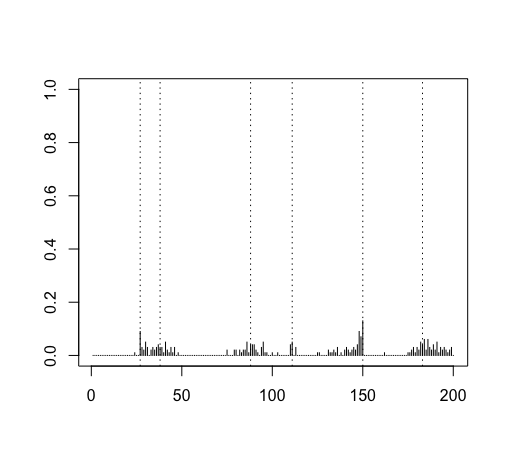}};
\node at (5,-8) {\includegraphics[height=4cm,width=5cm]{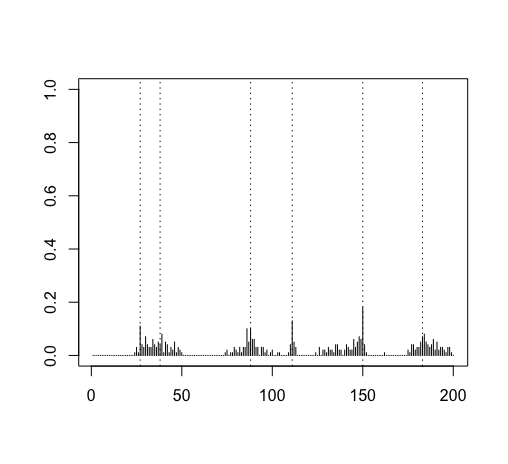}};
\node at (10,-8) {\includegraphics[height=4cm,width=5cm]{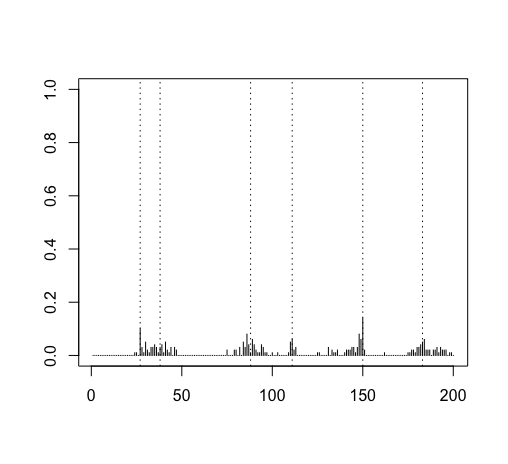}};
\node at (-2.6,1.3) {\footnotesize{$(a)$}};
\node at (-2.6,-2.7) {\footnotesize{$(b)$}};
\node at (-2.6,-6.7) {\footnotesize{$(c)$}};
 \end{tikzpicture}
\caption{Frequencies of each possible breakpoint for MHomo when the number of segments is selected with the criteria mBIC (left), Lav (middle) and BM2 (left), with $n=200$. The value of  $\sigma_2^\star$ is fixed to $0.1$ (a), $0.5$ (b) and  $1.5$ (c). The dotted lines correspond to the true breakpoint locations.} \label{fig:Hist_rupt_mod1}
\end{figure}
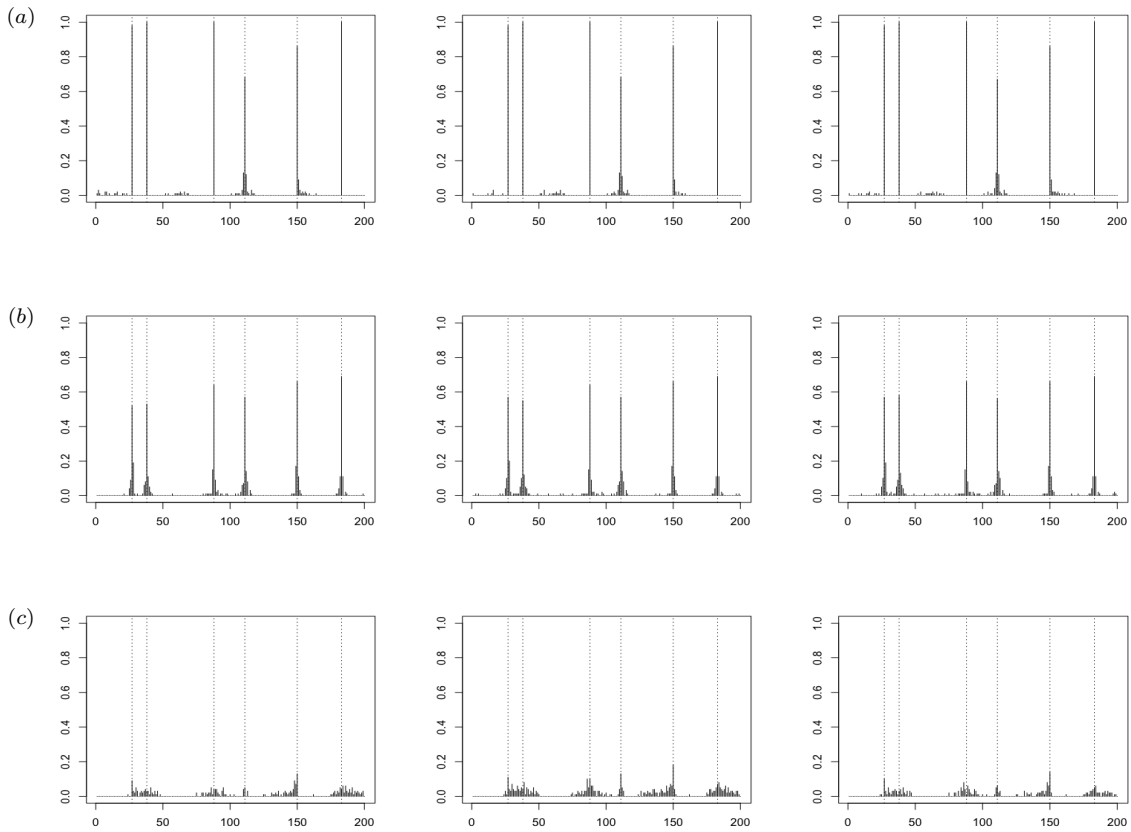
\begin{figure}[h!]
\centering
\begin{tikzpicture}
\node at (0,0) {\includegraphics[height=4cm,width=5cm]{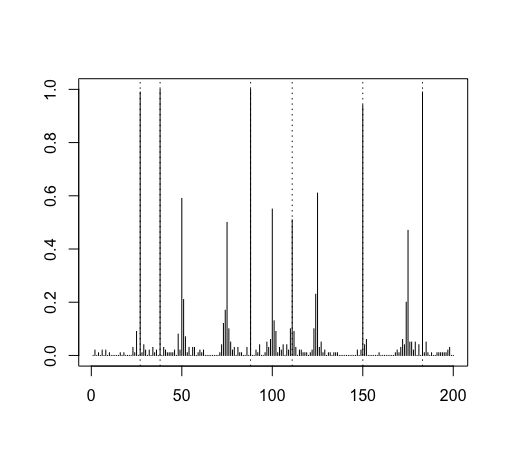}};
\node at (5,0) {\includegraphics[height=4cm,width=5cm]{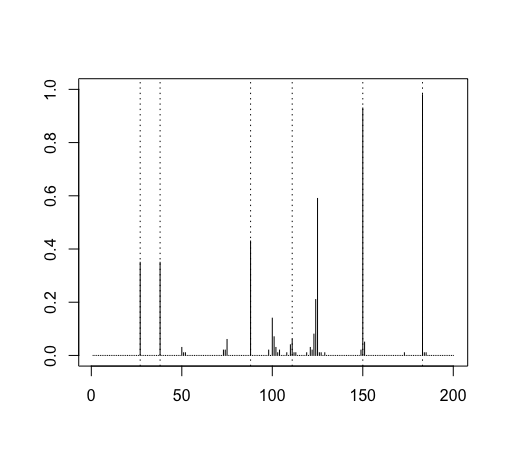}};
\node at (10,0) {\includegraphics[height=4cm,width=5cm]{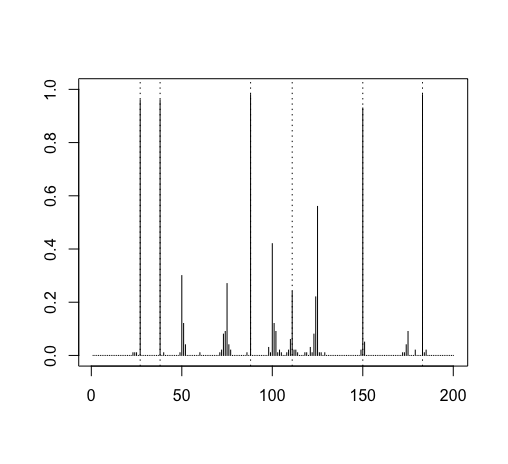}};
\node at (0,-4) {\includegraphics[height=4cm,width=5cm]{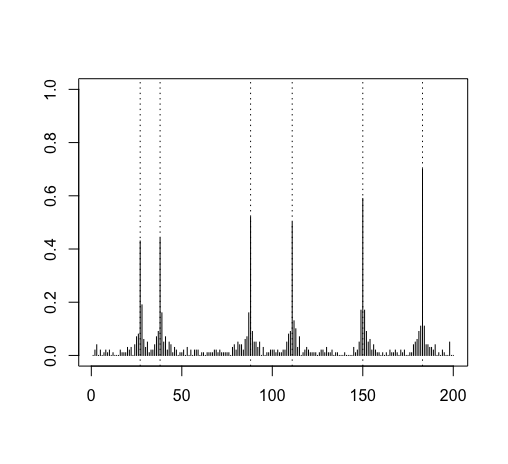}};
\node at (5,-4) {\includegraphics[height=4cm,width=5cm]{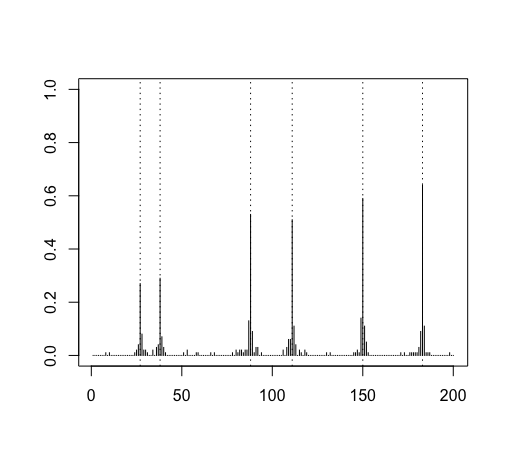}};
\node at (10,-4) {\includegraphics[height=4cm,width=5cm]{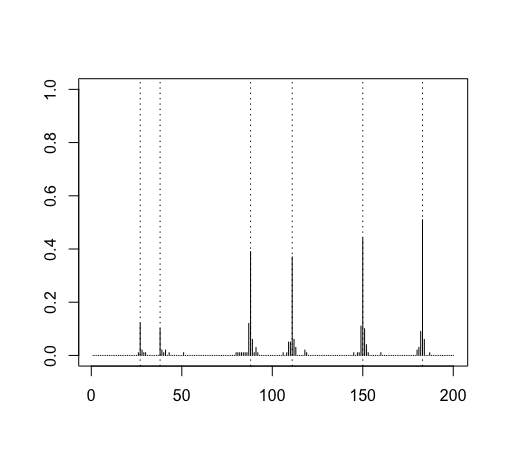}};
\node at (0,-8) {\includegraphics[height=4cm,width=5cm]{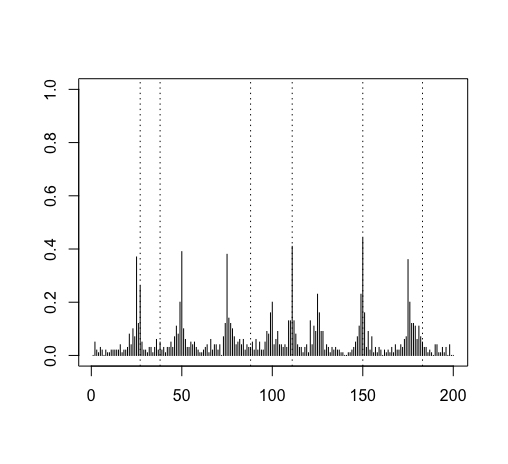}};
\node at (5,-8) {\includegraphics[height=4cm,width=5cm]{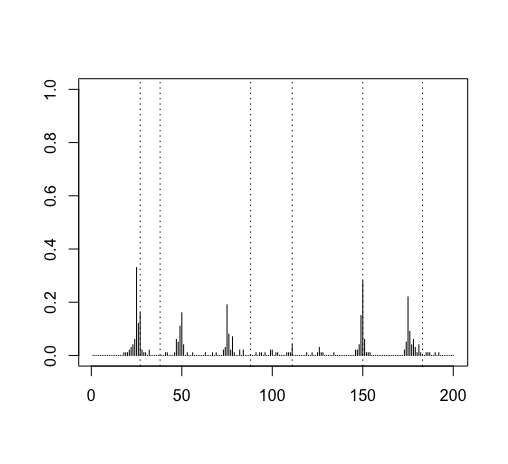}};
\node at (10,-8) {\includegraphics[height=4cm,width=5cm]{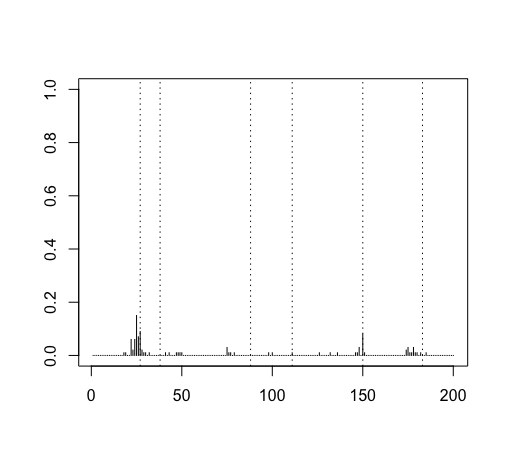}};
\node at (-2.6,1.3) {\footnotesize{$(a)$}};
\node at (-2.6,-2.7) {\footnotesize{$(b)$}};
\node at (-2.6,-6.7) {\footnotesize{$(c)$}};
 \end{tikzpicture}
\caption{Frequencies of each possible breakpoint for MHetero when the number of segments is selected with the criteria mBIC (left), Lav (middle) and BM2 (left), with $n=200$. The value of  $\sigma_2^\star$ is fixed to $0.1$ (a), $0.5$ (b) and  $1.5$ (c). The dotted lines correspond to the true breakpoint locations and the changes of variances are fixed at locations $25, 50, 75, 100, 125, 150, 175$.} \label{fig:Hist_rupt_mod2}
\end{figure}

\section{Application to GNSS-derived Integrated Water Vapour series} \label{sec:application}

\paragraph{\bf {Context and data description.}} The GNSS-derived IWV series are used to study and verify climate model predictions of atmospheric water vapour trends and variability connected to climate change (global warming) \cite{parracho2018global}. The biases induced by the abrupt changes are small and thus difficult to distinguish from the natural climate variation of the measured IWV signal \cite{ning2016uncertainty}. The most commonly used approach is the relative abrupt change detection which compares the candidate series to one or several reference series (e.g. from nearby stations) which are assumed to contain nearly the same climate signal \cite{lindau2013multiple,CM2004}. In the case of our application, the stations in the global GNSS network are usually too far from each other to remove completely the climate signal in the differences. Instead, we extract the reference time series for each station from a gridded global atmospheric model reanalysis. In this work, we use the ECMWF reanalysis, ERA-Interim (ERAI) \cite{dee2011era}. The considered series are thus the differences between the GNSS-IWV signal and the ERA-Interim one, resulting in so called GNSS-ERAI series.

In this application we consider the two stations SYOG (Syowa, Antarctica) and ONSA (Onsala, Sweden) contributing to the International GNSS Service (IGS) network of continuously operating reference stations (www.igs.org). The IWV data retrieved from these GPS measurements are described in \cite{parracho2018global}. In the present work, the IWV data series are used with daily time sampling. The equipment changes are available from the so-called IGS sitelogs and are given in Table \ref{tab:known-changes}.  \\

\paragraph{\bf {Model \eqref{model:m1} for this application.}}
For these series, the variance time-intervals correspond to the different months, i.e. 
\begin{description}
\item[$\star$] $j=\text{month}$, $J=12$ and each interval $I^{\text{var}}_\text{month}$ is the union of several intervals among the considered years, $I^{\text{var}}_\text{month} =\bigcup_{\text{year}} I^{\text{var}}_{\text{year},\text{month}}$ where $I^{\text{var}}_{\text{year},\text{month}}$ is the time-interval of the month 'month' of the year 'year'.
\item[$\star$] $\sigma^2_j$ is estimated by $Q_{n}(\xbf_\text{month})$ with $\xbf_\text{month}=((y_{t+1}-y_t)_{date(t) \ \text{and} \ date(t+1) \in \text{month}}^{\text{year}})_{\text{year}}$, i.e. the differentiated series of the considered month of all the years, and where $Q_{n}$ is defined by \eqref{eq:esti_sigma}.
\end{description}

\paragraph{\bf {Results.}}

\begin{figure}[h!]
\centering
\begin{tikzpicture}
\node at (0,6) {\includegraphics[height=6cm,width=8cm]{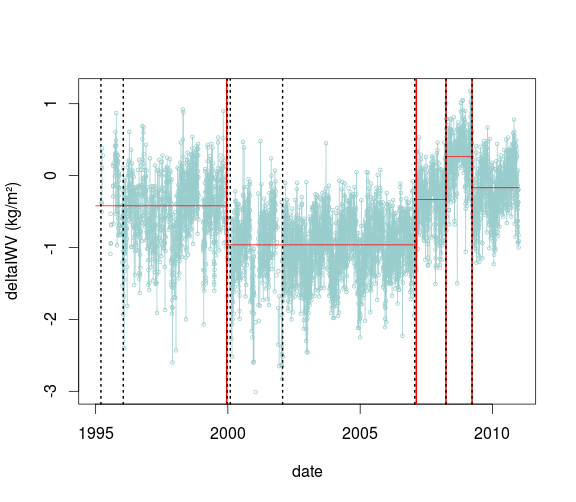}};
\node at (0,0) {\includegraphics[height=6cm,width=8cm]{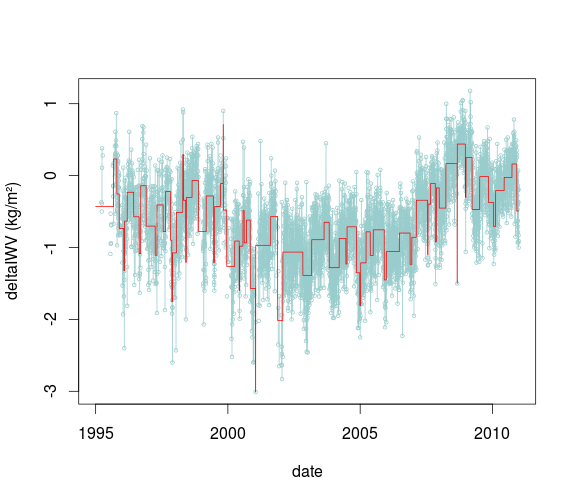}};
\node at (0,-6) {\includegraphics[height=6cm,width=8cm]{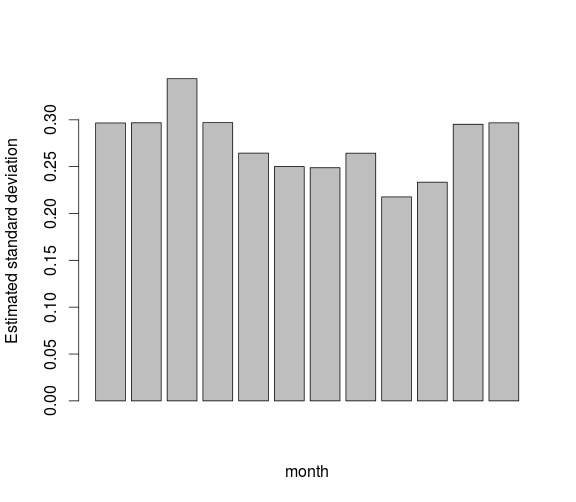}};
\end{tikzpicture}
\caption{Results for the series SYOG: the estimated breakpoint with $\hat{K}=5$ (top), the series with the estimated mean with $\hat{K}=82$ (middle) and the estimated standard deviation for each month (bottom). Solid lines (in red): the estimated breakpoints and the fitted expectation. Dashed lines (in black): known equipment changes (see Table \ref{tab:known-changes}).} \label{fig:SYOG}
\end{figure}

For the SYOG series, all the criteria select four breakpoins, except for mBIC that selects $81$ ones. The results are plotted in Figure \ref{fig:SYOG}. All the four breakpoints correspond (exactly for dates  2008-03-31 and 2009-03-26 and are close for dates 1999-12-16 and 2007-02-15) to known equipment changes (the dashed lines (in black)). This segmentation is also obtained by both the models MHomo and MHetero with BM2. This can be explained by the fact that the monthly variances are quite similar (see the estimated standard deviation of each month Figure \ref{fig:SYOG} (middle)). \\

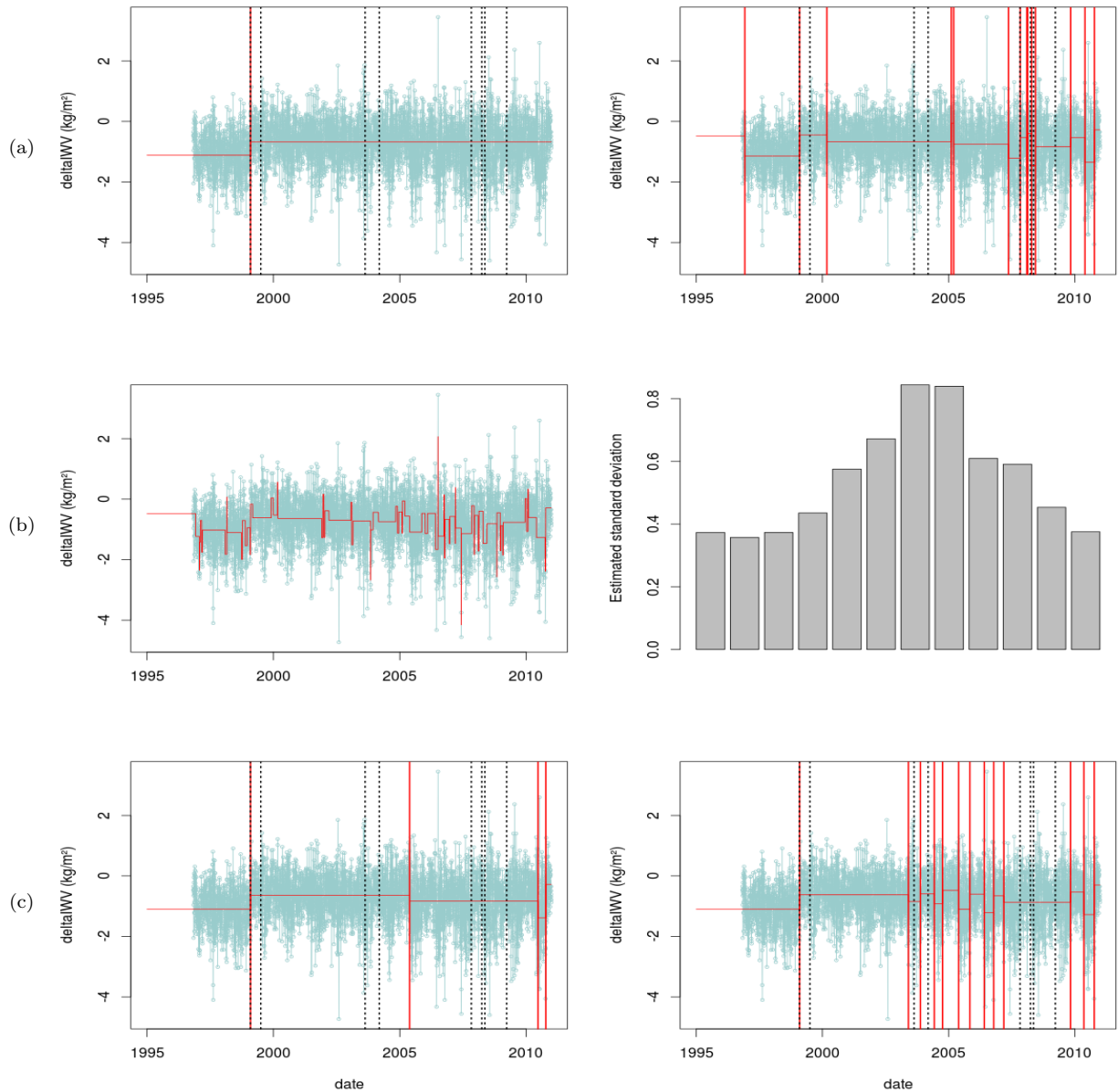
\begin{figure}[h!]
\centering
\begin{tikzpicture}
\node at (0,0) {\includegraphics[height=6cm,width=8cm]{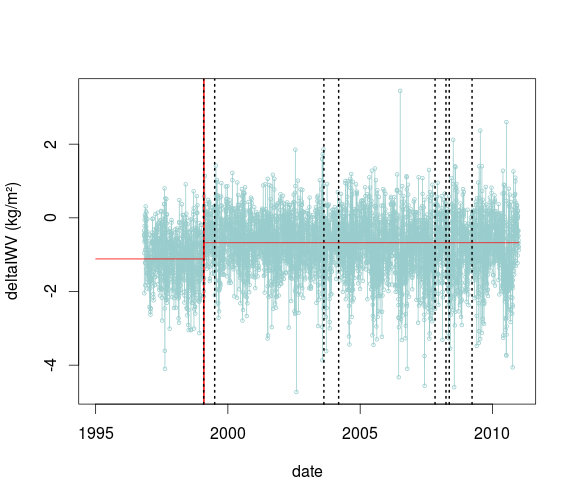}};
\node at (8,0) {\includegraphics[height=6cm,width=8cm]{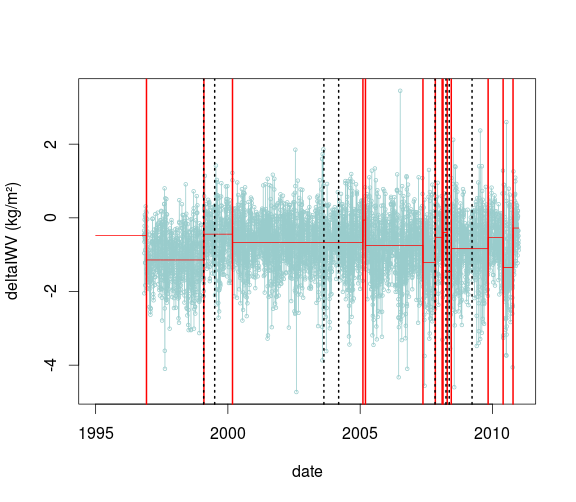}};
\node at (0,-5.5) {\includegraphics[height=6cm,width=8cm]{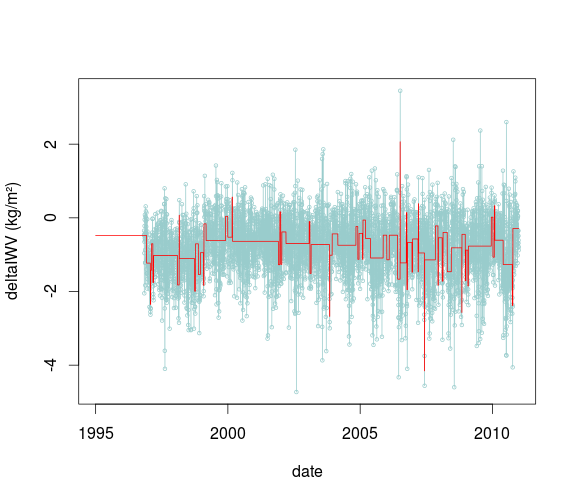}};
\node at (8,-5.5) {\includegraphics[height=6cm,width=8cm]{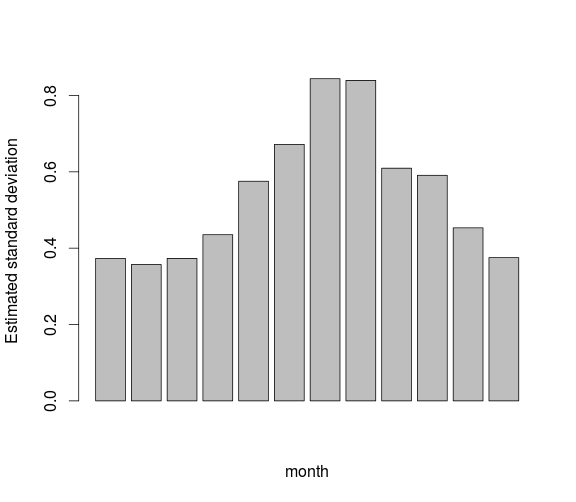}};
\node at (0,-11) {\includegraphics[height=6cm,width=8cm]{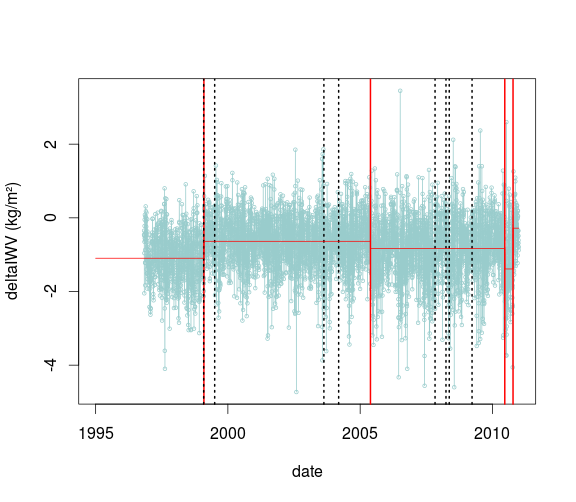}};
\node at (8,-11) {\includegraphics[height=6cm,width=8cm]{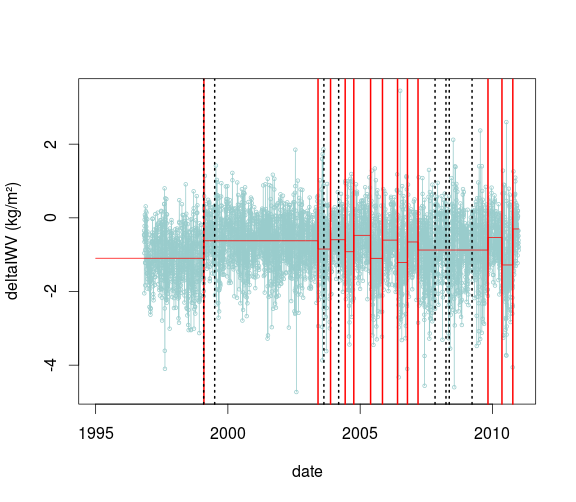}};
\node at (-4.5,0) {\footnotesize{(a)}};
\node at (-4.5,-5.5) {\footnotesize{(b)}};
\node at (-4.5,-11) {\footnotesize{(c)}};
\end{tikzpicture}
\caption{Results for the series ONSA. (a) the estimated breakpoints with $\hat{K}=2$ (left: Lav and BM2) and $\hat{K}=15$ (right: BM1). (b) the estimated mean with $\hat{K}=74$ (left: mBIC) and the estimated standard deviation for each month (right). (c) the estimated breakpoints obtained with model MHomo (left: $\hat{K}=5$ with Lav) and with model MHetero (left: $\hat{K}=14$ with Lav). Solid lines (in red): the estimated breakpoints and the fitted expectation. Dashed lines (in black): known equipment changes (see Table \ref{tab:known-changes}). } \label{fig:ONSA}
\end{figure}

The results for the series ONSA are given in Figure \ref{fig:ONSA}. The criteria select different number of segments: $\hat{K}=2$ for Lav and BM2, $\hat{K}=15$ for BM1 and $\hat{K}=76$ for mBIC. The big abrupt change at date 1999-02-04 is always detected and is associated to a change in receiver, antenna and radome. When $\hat{K}=15$, only one break corresponds to a known change and two others are close. Contrary to the previous series, the estimated monthly variances are different (higher in summer) resulting in a different segmentation for models MHomo and MHetero (see Figure \ref{fig:ONSA} (c) where the criterion Lav is considered). The breakpoint at date 1999-02-04 is detected with the both. However we observe an overestimation of the number of breakpoints and the estimated breakpoints are clearly not linked to known equipment changes. Note that all equipment changes do not impact the time series \cite{ning2016uncertainty}. \\    

For both series, we observe an overestimation of the number of segments when using the mBIC criterion. By looking to the estimated means (Figure \ref{fig:SYOG} (middle) and Figure \ref{fig:ONSA} b-left), this overestimation links to the detection of outliers and seems to capture a periodic signal. This latter point can be due to the fact that a periodic tendency remains despite the correction by the ERAI model \cite{parracho2018global}.

\begin{table}
\begin{center}
{\small
\begin{tabular}{lp{.3\textwidth}}
Series & known changes  \\
\hline
SYOG & 1995-03-15 (RA) \\
& 1996-01-17 (R) \\
& 1999-12-24 (R) \\
& 2000-02-03 (R) \\
& 2002-01-26 (R) \\
& 2007-01-25 (R) \\
& 2008-03-31 (P) \\
& 2009-03-26 (P) \\
\hline
ONSA & 1999-02-01 (RAD) \\
& 1999-07-03 (R) \\
&  2003-08-19 (R)\\
& 2004-03-10  (R) \\
& 2007-11-01 (R) \\
&  2008-03-31 (P) \\
&  2008-05-15 (R) \\
&  2009-03-26 (P)  \\
\hline
\end{tabular}
}
\end{center}
\caption{Known changes in the two considered series. All changes corresponds to a change of receiver (R), antenna (A), radome (D), or processing (P). RA and RAD indicate combined changes.}
\label{tab:known-changes}
\end{table}

\bibliographystyle{bmc_article}
\bibliography{biblio}
\end{document}